\input iopppt.tex
\pptstyle
\letter{The stress--energy operator}

\author{ Adam D Helfer}

\address{Department of Mathematics, Mathematical Sciences Building, 
University of Missouri, Columbia, Missouri 65211, U.S.A.}

\jl{6}

\beginabstract We compute the stress--energy operator for a scalar
linear quantum field in curved space--time, modulo c--numbers.  For the
associated Hamiltonian operators, even those generating evolution along
timelike vector fields, we find that in general on locally Fock--like
(`Hadamard') representations:  (a) The Hamiltonians cannot be
self--adjoint operators; (b) The automorphisms of the field algebra
generated by the evolution cannot be unitarily implemented; (c) The
expectation values of the Hamiltonians are well--defined on a dense
family of states; but (d) These expectation values are unbounded below,
even for evolution along future--directed timelike vector fields and 
even on Hadamard states.
These are all local, ultraviolet, effects.  
\endabstract


\date

\font\tenbm=msbm10
\font\sevenbm=msbm7
\newfam\msyfam
\textfont\msyfam=\tenbm
\scriptfont\msyfam=\sevenbm
\def\Bbb#1{{\fam\msyfam #1}}
\def\R{{\Bbb R}}
\def\C{{\Bbb C}}
\def\wp{{\widehat\phi}}

\def\wH{{\widehat H}}
\def\wh{{\widehat H}}
\def\whT{{\widehat T}}
\def\sym{\mathop{\rm sym}}
\newcount\EEK
\EEK=0
\def\eek{\global\advance\EEK by 1\eqno(\the\EEK )}

\nosections 
In a relativistic quantum field theory, two sorts of
evolution are important:  that of the state vectors; and that of the
field operators.  The state vectors are required to evolve by unitary
motions of the Hilbert space (except when reduction occurs) in order
that probability be conserved.  On the other hand, the evolution of the
field operators is governed by relativistic field equations on
space--time, and need not arise from unitary motions.

To spell out what this means, we recall that a relativistic field
theory is constructed by first forming an algebra ${\cal A}$ of fields,
encoding the canonical commutation relations.  This algebra is abstract
in that it is not (yet) represented as a collection of operators on a
physical Hilbert space; the choice of such a representation is the
second step in the construction.  Now evolution in space--time is
represented by flow along a vector field, and this flow induces a
family of automorphisms of ${\cal A}$.  But these automorphisms need
not arise from unitary motions on a given representation of ${\cal
A}$.  We show here that this failure occurs in general when (as is
usual) the representation is taken to be `Hadamard' --- that is, to
have the same leading short--distance behavior as the Fock
representation in Minkowski space.

The phenomenon of non--unitarily implementable evolution has been known
for some time, but has generally been thought possible only when
unaccounted--for massless radiation might escape to infinity (the
`infrared catastrophe'; see Bjorken and Drell 1965, Jauch and Rohrlich
1975, Ashtekar and Narain 1981).  Here it occurs as a local
(ultraviolet) effect.  Since the evolution is supposed to be generated
by Hamiltonian operators, it is perhaps not surprising that the lack of
unitary implementability is tied to a lack of existence of fully
satisfactory Hamiltonians.  These cannot exist as operators on
reasonable domains in the physical Hilbert space.  Their expectation
values are well--defined on a dense family of states, but generically
these expectation values are unbounded below.

The present results should be contrasted with those of Brunetti et al
(1995).  These authors proved that the stress--energy operator exists
\it as an operator--valued distribution \rm on Hadamard
representations.  This means that $\int \whT _{ab} f^{ab}\, \d{\rm
vol}$ exists as a self--adjoint operator when $f^{ab}$ is taken to be a
compactly supported test--function, smooth on space--time.  However,
the Hamiltonian operators are given by integrating $\whT _{ab}$ over
Cauchy surfaces (equivalently, by taking $f^{ab}$ to be a kind of
distribution), and so the work of these authors is not in conflict with
the present result.  Comments about the physical significance of the
two results will be found at the end of this letter.

The arguments of Brunetti et al are general but not explicitly
computational, and they establish the existence of the stress--energy
defined by the point--splitting procedure (cf. Wald 1994).  It will
turn out that our stress--energy's expectations verify Wald's key
axioms (his (1) and (2)), and then Wald's work shows that our
stress--energy is the same as that of Brunetti et al (modulo
c--numbers).

This letter is given to outlining the arguments in a simple case, that
of an uncharged spinless field of mass $m$ in a curved space--time
$(M,g_{ab})$.  Similar effects occur for fields responding to scalar
and electromagnetic potentials in Minkowski space; these will be
treated elsewhere, along with many mathematical details.  We use the
by--now standard symplectic quantization, which is treated well in
Wald's (1994) book; see also Ashtekar and Magnon--Ashtekar (1980),
Birrell and Davies (1982), DeWitt (1983), Fulling (1989).  See Reed and
Simon (1972--5) for general operator theory and Treves (1980)  for
pseudodifferential operators.  Our conventions for space--time
quantities are those of Penrose and Rindler (1984--6).  We set $c=\hbar
=1$.

The field equation is
$$(\nabla _a\nabla ^a+m^2)\phi =0\, .\eek$$\xdef\feq{\the\EEK}%
We assume the space--time is oriented, time--oriented and globally
hyperbolic.  We also take the Cauchy surfaces to be compact;  
this is done only for simplicity:  it avoids some operator--theoretic 
technicalities and rules out all infrared effects.  We let $\Gamma$ 
be the space of smooth classical solutions to (\feq ); it is
equipped with a symplectic form $\omega (\phi ,\psi )=
\int _\Sigma \bigl( \psi \, {}^*\d\phi -\phi \, {}^*\d\psi\bigr)$.  
See Wald (1994) for a precise definition of ${\cal A}$.

The representations 
of interest are mathematically analogous to the
Fock representation.  Let $J:\Gamma\to\Gamma$ be a map preserving
$\omega$, satisfying $J^2=-1$ and such that
$$\omega (\phi ,J\phi )\eek$$\xdef\rnorm{\the\EEK}%
is positive--definite.  (In Minkowski space, one chooses $J\phi =\i
(\phi _+-\phi _-)$, where $\phi _\pm$ are the positive-- and
negative--frequency parts of $\phi$.)
Then $J$ makes $\Gamma$ (completed with respect to
the norm (\rnorm )) a complex Hilbert space $\Gamma (J)$, 
which is the analog of the
one--particle Hilbert space,\footnote{\dag}{This analogy is in general
only mathematical.  The identification of which mathematical structures
correspond to physical one--particle states is involved and to some
degree ill--defined.  See Wald (1994) and references therein.}
and the full Hilbert space may be
constructed from this as usual.  In fact, it is a little more 
convenient to start with $\Gamma (-J)$; then the full Hilbert space 
${\cal H}(J)$ may be identified with a space of holomorphic (rather 
than antiholomorphic) functions on $\Gamma (J)$, the elements of the
symmetric tensor powers of $\Gamma (-J)$ forming the coefficients of 
the power--series expansion.  On such a function $\Phi (\phi )$, the
action of the field
operators is determined by
$\omega (\alpha ,\wp )\Phi (\phi )=\omega (\alpha _-,\phi )\Phi
  +\i\nabla _{\alpha _+}\Phi$,
where $\alpha _\pm$ are the $J$--linear and $J$--antilinear parts of
the classical solution $\alpha$.  It should be noted that ${\cal H}(J)$
contains only the holomorphic functions whose norm (in the appropriate
sense) is finite.

The choice of $J$ is important.  In general, different choices
will not lead to unitarily equivalent representations; one has unitary
equivalence iff their difference is Hilbert--Schmidt, that is, iff
${\rm tr} (J_1-J_2)^2<\infty$.  Kay and Wald (1991) 
have shown that there is a
choice of $J$ such that the two--point functions $\langle \wp (x)\wp
(y)\rangle$ of the $J$--`vacuum'
have the same asymptotic form as $x\to y$ (in geodesic 
normal coordinates) as for the Fock representation, and that 
(in the case of compact Cauchy surfaces) this choice is
unique up to unitary equivalence.  Such a representation is called
\it Hadamard.  \rm  For the remainder of this paper, we assume the
physical Hilbert space is a Hadamard representation.

We now take up the problem of evolution.
The algebra ${\cal A}$ is generated by the Cauchy
data for $\wp$ on any Cauchy surface.  Flowing along a vector field 
$\xi ^a$
in $M$ induces a canonical transformation of Cauchy data, which
generates an automorphism of ${\cal A}$.  Our first result is
that in general this automorphism need not be induced by a unitary 
map on ${\cal H}(J)$.  In other words, the automorphism
may be regarded as determining a \it deformation \rm of the
representation.  

Consider a space--time with $M=\{ (t,x)\mid t\in \R,$ $x\in\Sigma\}$, 
where $t$ is a global time function and for
$t\leq t_-$ (resp., $t\geq t_+>t_-$) the metric is static of the form
$$\d t^2-h^\pm _{ab}\d x^a\d x^b\, ,\eek$$
where $h^\pm _{ab}$ is a $t$--independent positive--definite
three--metric on $\Sigma$.
The vector field $\xi$ is taken to be $\partial /\partial t$, and the 
Cauchy surfaces to be surfaces of constant $t$.
We shall write $\Gamma _\Sigma =\{ (\phi ,\pi )\in C^\infty (\Sigma
)\times C^\infty (\Sigma,\wedge ^3 )\}$ for the space of Cauchy data
(using the `momentum' $\pi ={}^*\d\phi$ pulled back to $\Sigma$ in 
place of the normal derivative).  Then we have isomorphisms 
$i_t:\Gamma _\Sigma\to\Gamma$ embedding these as classical solutions 
with data given at time $t$. 

Suppose that evolution were unitarily implementable.  
Let $J$ be
a Hadamard complex structure on $\Gamma$.  An allowable
choice of $J$ in the regime $t<t_-$ (resp., $t>t_+$), in block form
with respect to the decomposition above, is
$i_{t_\pm}J_\pm{}i_{t_\pm}^{-1}$, where
$$J_\pm =\left[\matrix{0&\bigl( m^2-\Delta _\pm \bigr) ^{-1/2}
    (\d{\rm vol}(h_\pm ))^{-1}\cr
  -(\d{\rm vol}(h_\pm ))
  \bigl( m^2-\Delta _\pm \bigr) ^{1/2}&0\cr}\right]\,
,\eek$$\xdef\adJ{\the\EEK}%
and $\Delta _\pm$ is the Laplacian associated with $h^\pm
_{ab}$.
Evolution from $t_-$ to $t_+$ is given by the automorphism
$g=i_{t_+}i_{t_-}^{-1}$ of $\Gamma$.  Our hypothesis is that this is
unitarily implementable, and this implies that the complex structure
$gi_{t_-}J_-{}i_{t_-}^{-1}g^{-1}$ is Hadamard.  But 
$gi_{t_-}J_-{}i_{t_-}^{-1}g^{-1}=i_{t_+}J_-{}i_{t_+}^{-1}$.
Therefore the structures
$$i_{t_-}J_-{}i_{t_-}^{-1}\, ,\ i_{t_+}J_+{}i_{t_+}^{-1}\, ,\ 
 i_{t_+}J_-{}i_{t_+}^{-1}\eek$$
are all Hadamard, and by equivalence of Hadamard structures, the
difference between any two must be Hilbert--Schmidt.  Taking the
difference of the last two, and conjugating by $i_{t_+}$ (which
preserves the Hilbert--Schmidt class) implies that $J_--J_+$ must be
Hilbert--Schmidt.  However, it is easy to see that this fails in
general.  (For example, if $h^-_{ab}=Ch^+_{ab}$ for a constant $C$, 
then ${\rm tr}(J_--J_+)^2$ is explicitly calculable in terms of the
eigenvalues of the spatial Laplacian.)  
This is a contradiction, and so the evolution
cannot be unitarily implemented.

The foregoing is a rigorous proof that 
a certain class of space--times, vector
fields and Cauchy surfaces has
non--unitarily implementable evolution.  But one expects the result to
hold generically, since ${\rm tr} (J-gJg^{-1})^2$ can be
expressed as an integral of products of Green's functions and kernels
for the two--point function, so the fact that it diverges should be
stable.

Although the evolution is not unitarily implementable, one can still
search for a generator, that is, an operator $\wh (\xi ,\Sigma )$ such
that
$$[\wh (\xi ,\Sigma ),\wp ]=-\i\xi ^a\nabla _a\wp\quad
\hbox{ on }\Sigma\eek$$\xdef\cttr{\the\EEK}%
(and the usual formula for $[\wh (\xi ,\Sigma ),\nabla _{\rm n}\wp ]$, 
with $\nabla _{\rm n}$ the normal derivative).
Such an
operator does exist, although in a limited sense.
We adopt a standard index notation for the complex vector space
$\Gamma (J)$; we write $\phi ^\alpha$ for an arbitrary element of this
space, and $\eta _{\overline{\alpha} \alpha}$ for the inner product. 
Then the classical Hamiltonian operator $h(\xi ,\Sigma )$ on $\Gamma$
may be decomposed into a $J$--linear part $iB^\alpha{}_\beta$ and an
antilinear part $C^{\overline\alpha}{}_\beta$.  We set
$A_{\alpha\beta} =(-\i /4)
  \eta _{\overline\alpha \alpha}C^{\overline\alpha}{}_\beta$.  
Then formally 
$$\wH (\xi ,\Sigma )=A_{\alpha\beta}\phi ^\alpha\phi ^\beta
+B^\beta{}_\alpha \phi ^\alpha {{\partial}\over{\partial\phi ^\beta}}
  +{\overline A}^{\alpha\beta}{{\partial}\over{\partial\phi ^\alpha}}
  {{\partial}\over{\partial\phi ^\beta}}\eek$$\xdef\ham{\the\EEK}%
generates evolution.
Note that
$\wh (\xi ,\Sigma )$
is equal to the
classical Hamiltonian function with $\wp$ replacing $\phi$ and
normal--ordered with respect to the senses of
creation and annihilation defined by $J$.  Therefore it arises 
from a
stress--energy operator $\whT _{ab}$ which is the $J$--normal ordered
transcription of the classical stress--energy.
It can be shown from this that the expectation values $\langle\Phi |
{\widehat T}_{ab}|\Phi\rangle$ verify Wald's (1994) axioms.
One sees that $\wH (\xi
,\Sigma )$ is determined modulo c--numbers by the requirement
that it generate evolution.  

For stronger results, we need a usable formula for $J$. 
A straightforward lengthy calculation shows that in normal coordinates
(for the induced metric on $\Sigma$), the symbol of $J$ is given by
$$\sym J(\zeta )=\left[\matrix{\sym\alpha &|\zeta |^{-1}\cr -|\zeta |&
    -\sym\alpha\cr}
    \right] 
    +\left[\matrix{O\bigl( |\zeta |^{-2}\bigr)
    &O\bigl( |\zeta |^{-3}\log |\zeta |\bigr)\cr
  O\bigl( |\zeta |^{-1}\log |\zeta |\bigr)
    &O\bigl( |\zeta |^{-2}\bigr)\cr}\right]\, ,\eek
$$
where $2\sym\alpha =
|\zeta |^{-1}\pi _a{}^a -|\zeta |^{-3}\pi ^{ab}\zeta _a\zeta _b$
and $\pi _{ab}$ is the second fundamental form of $\Sigma$.
(The operator $J$ is obtained to corresponding accuracy by putting
$-\i\nabla _a$ for $\zeta _a$.  See Treves 1980.)  With this
in hand, it is also clearest to invoke the isomorphism of $\Gamma (J)$
with the $+\i$ eigenspace of $J$ in $\Gamma\otimes \C$ to identify 
$\phi ^\alpha$ with the complex wave--function $\phi _{\rm cx}$, 
defined as the configuration
value of $(1/2)(1-\i J)\phi$.\footnote{\dag}{It is
possible for the projection onto the
configuration values to have a finite--dimensional kernel.  However,
this does not happen generically, and in any event it is not hard to
modify what follows to apply to this case.}
Then with
$s=\sqrt{m^2-\Delta }$,
in the case $\xi ^a=\beta t^a$, where $t^a$ is the unit normal to
$\Sigma$, we find
$$\left.\eqalign{
  \eta _{\overline\alpha \alpha}{\overline\phi}^{\overline\alpha}\phi
^\alpha &=\int \overline{\phi _{\rm cx}}
   \bigl( s+\hbox{l.o.t.}\bigr)\phi _{\rm cx}\d\Sigma\cr
  A_{\alpha\beta}\phi ^\alpha\phi ^\beta
    &=(\i /4)\int \bigl(\alpha\beta s^2+\hbox{l.o.t.}\bigr)
    \phi _{\rm cx}\cdot\phi _{\rm cx}\d\Sigma\cr
  B\phi _{\rm cx}&=\bigl(\beta s+\hbox{l.o.t.}\bigr)\phi _{\rm cx}\cr}
\right\}\, ,\eek$$\xdef\sleek{\the\EEK}%
where `l.o.t.' means `lower--order terms' in the sense of
pseudo--differential operators (Treves 1980).

Remarks:  
(a) A computation like that of (\sleek ), but keeping all
terms shows that $B^\alpha{}_\beta$ is formally self--adjoint, and that  
for $\xi ^a$ everywhere timelike, the operator $B^\alpha{}_\beta$ is 
a real elliptic pseudodifferential operator and has a canonical 
self--adjoint extension.
(b) From (\sleek ), we see $A_{\alpha\beta}$ is a bounded 
symmetric form, but that  
generically
$A_{\alpha\beta}{\overline A}^{\alpha\beta}=\infty$, so 
$\wh (\xi ,\Sigma )$ cannot be an operator from any domain containing 
the constant functions to ${\cal H}(J)$.
(c) However $\wh (\xi ,\Sigma 
)$ does exist as a quadratic form on, for example, the
space of polynomials each of whose terms is a tensor of finite rank 
over the dual of the domain of $B^\beta{}_\alpha$.  Such a domain is 
dense in ${\cal H}(J)$.  Then $\wh (\xi ,\Sigma )$ can be given 
meaning as an operator from this domain to a certain space of 
polynomials, not elements in ${\cal H}(J)$.  In this sense  
(\cttr ) holds. 

Our next result, perhaps the most disturbing, is this:  that even for
$\xi ^a$ timelike future--pointing, generically the form $\wh (\xi ,
\Sigma  )$ is
unbounded below.  
Take $\xi ^a$ to be
the unit normal to $\Sigma$, for simplicity.  Then using (\sleek ) one
can show that states of the form
$\Phi (\phi )=1+b_{\alpha\beta}\phi ^\alpha\phi ^\beta$ with
$$b_{\alpha\beta}\phi ^\alpha\phi ^\beta
  =-kA_{\alpha\beta}\phi ^\alpha (B^\nu\phi )^\beta
\, ,\eek$$\xdef\slaak{\the\EEK}%
with
$k$ a positive constant, for $-3<\nu <-2$, have finite norm but
$\langle\Phi |\wh (\xi ,\Sigma )|\Phi\rangle =-\infty$. 
Now $b_{\alpha\beta}\phi ^\alpha\phi ^\beta$
is given by integrating the Cauchy data for $\phi ^\alpha$ against a
certain distributional kernel.  We may approximate this distribution by
a sequence of smooth kernels, giving a sequence of states $\Phi _j(\phi
)$.  Each of these is a \it Hadamard state, \rm that is, has two--point
functions with singular parts
given by the classical Hadamard expansion (Wald 1994).  
The states $\Phi _j$ are uniformly bounded in norm and have
$\langle\Phi _j|\widehat H (\xi ,\Sigma )|\Phi _j
\rangle\to -\infty$.

Finally, for $\Sigma$ a
surface of constant time in Minkowski space, if $\xi ^a$ is not a
Killing vector, similar calculations show
that still the Hamiltonian fails to exist
as a self--adjoint operator, and still is unbounded below as a form.
(Compare Epstein et al 1965.) 
Thus what distinguishes special from general relativity is the \it
ability \rm to find $\Sigma$ and $\xi ^a$ such that the Hamiltonians
have satisfactory properties.  This ability depends on the
existence of hypersurfaces with symmetries.  (The leading pathogenic
terms cancel iff there is a solution to Killing's equation projected to
$\Sigma$.)

We now discuss the consequences of these results.  
The central question they raise is, How is energy
exchanged between the quantum field and gravitational (or other)
fields?  

Insofar as it is possible to accept the external--field approximation as
a model of physics, and deferring the identification of a self--adjoint
energy operator,
the results uncovered here are simply curiosities:  they do
not affect the dynamics of the theory.  However, this
neglects interactions; the real world would be
described by the model subject to perturbations.  If 
$\langle\int{\widehat T}_{ab}\xi ^a\d\Sigma ^b\rangle$ were any sort of 
measure of the energy of a state, one
would expect the perturbations to lead to the emission of arbitrarily
large energies from the quantum field, an effect which is not observed.
(This concern is parallel to that which led to the second
quantization of the electron field, of course.)  Either one must explain
why perturbations do not have this effect, or one must abandon the
interpretation of the expectation value as a measure of energy.

The non--existence of the Hamiltonians as self--adjoint operators means
that, in arbitrarily weak gravitational fields, we do not have a
satisfactory definition within conventional quantum theory of the
energy as an observable.  This leads to two questions.  Is there a
satisfactory definition, perhaps of another nature?  And, If 
$\int \whT _{ab}\xi ^a\d\Sigma ^b$ is not an observable, 
what quantity is measured in experiments?

Some progress on these questions ought to be possible by making
explicit enough models of energy measurement.  Recall that
the problems uncovered here occur in Minkowski space, if the vector
field $\xi ^a$ is not Killing.  But a real experiment, which occupies a
finite space--time volume, cannot measure $\int {\widehat
T}_{ab}\xi ^a\d\Sigma ^b$ for a Killing field (since then
$\xi ^a$ is not compactly supported).  
What we measure in the laboratory is some
self--adjoint operator which approximates $\int {\widehat
T}_{ab}\xi ^a\d\Sigma ^b$.  A model of measurement should identify these
approximants, and lead to an understanding of what bounds, if any, there
are on observable energies.  (Compare Ford and Roman 1995.)

A more speculative possibility, which will be considered
elsewhere, is that measurement of energy effects a change of
representation, the new representation being determined so that the
Hamiltonian exists as a semibounded self--adjoint operator.

All that has been used so far is the external--field approximation. 
If we assume as well the semiclassical approximation,
$$G_{ab}=
  -8\pi G\langle \whT _{ab}\rangle\, ,\eek$$\xdef\scapp{\the\EEK}%
we have all the previous problems, but additionally
the negative--energy states give rise to arbitrarily strong
repulsive gravitational
fields.  
A possible remedy 
is to use a space--time averaged stress--energy
operator.  Indeed, probably many workers on
semiclassical gravity have regarded (\scapp ) as a short--hand,
the right--hand side being more properly replaced by a
space--time average.  Recent intriguing `quantum inequalities' suggest
that this might remove the negatively infinite energies (Ford and Roman
1995, see also Flanagan and Wald 1996).  In this context, the force of
the present results is that, since  the limit of the average over
smaller space--time volumes does not exist, there is no ideal value to
which averages tend, and
the details of the
averaging must be spelled out.  It seems worthwhile pointing out
that, to have reasonable causality properties, one would expect $\whT
_{ab}(x)$ to be replaced by an average over the causal past $J^-(x)$;
this would introduce a manifest time--asymmetry into the theory.

The most striking applications of quantum field theory in curved
space--time --- the prediction that black holes should lose mass by
Hawking radiation, and the inflationary cosmologies --- depend on
the semiclassical approximation.  We must regard these
as at least thrown into question by the present results.

\ack
I thank Robert Wald for useful comments, and for 
drawing the paper of Brunetti et al to my attention.
This work was supported in part by the
University of Missouri through a
Research Leave, and conducted in part at the Mathematical Sciences
Research Institute, Berkeley.

\references

\refjl{Ashtekar, A and Magnon--Ashtekar, A 1980}{\GRG}{12}{205--25}

\refbk{Ashtekar, A and Narain, K S 1981}{Infrared Problems and
Penrose's Null Infinity}{(Berlin:  International Conference on
Mathematical Physics}

\refbk{Birrell, N D and Davies, P C W 1982}{Quantum Fields in Curved
Space--Time}{(Cambridge:  University Press)}

\refbk{Bjorken, J D and Drell, S D 1965}{Relativistic Quantum Fields}%
{(New York: McGraw--Hill)}

\refbk{Brunetti, R, Fredenhagen, K and Koehler, M 1995}{The microlocal 
spectrum condition and Wick polynomials of free fields on curved
spacetimes}{(gr-qc/9510056)}

\refbk{DeWitt, B S 1983}{{\rm in} Relativit\'e, groupes et toplogie II,
\rm eds. B. S. DeWitt and R. Stora}{(Amsterdam:  Elsevier)}

\refjl{Epstein, H, Glaser, V and Jaffe, A 1965}{\NC}{36}{1016--22}

\refbk{Flanagan, E and Wald, R M 1996}{Does backreaction enforce the
averaged null energy condition in semiclassical gravity?}{(gr-qc
9602052)}

\refjl{Ford, L H and Roman, T A 1995}{\PR}{D51}{4277--86}

\refbk{Fulling, S A 1989}{Aspects of Quantum Field Theory in Curved 
Space--Time}{(Cambridge: University Press)}

\refbk{Jauch, J M and Rohrlich, F 1975}{The Theory of Photons and
Electrons}{(New York:  Springer)}

\refjl{Kay, B S and Wald, R M 1991}{\it Phys. Rep.}{207}{49--136}

\refbk{Penrose R and Rindler W 1984--6}{Spinors and
Space--Time}{(Cambridge:  University Press)}

\refbk{Reed, M and Simon, B 1972--75}{Methods of Modern Mathematical
Physics, volumes I and II}{(New York:  Academic Press)}

\refbk{Treves, F 1980}{Introduction to Pseudodifferential and Fourier
Integral Operators}{(New York:  Plenum)}

\refbk{Wald, R M 1994}{Quantum Field Theory in Curved Spacetime and
Black Hole Thermodynamics}{(Chicago:  University Press)}

\bye